\documentclass[a4paper]{mn2e}
\usepackage{pslatex}
\ifx\pdftexversion\undefined
  \usepackage[dvips]{graphicx}
\else
  \usepackage[pdftex]{graphicx}
\fi
\usepackage[
pdfauthor={A.R. King, M.E. Beer, D.J. Rolfe, K. Schenker and
  J.M. Skipp},
pdftitle={The population of black widow pulsars},
pdfsubject={We discuss the formation and visibility of the observed
  Black Widow Pulsars},
pdfkeywords={binaries: close}
]{hyperref}

\def\msun{{\rm M_{\odot}}}

\date{Accepted 2005 January 27. Received 2005 January 26; in original form 2004 April 4}

\volume{000}

\pagerange{\pageref{firstpage}--\pageref{lastpage}} \pubyear{2005}

\begin{document}

\label{firstpage}

\title[The population of black widow pulsars]
{The population of black widow pulsars}

\author[A.\,R.~King et al.]{
A.\,R.~King\thanks{E-mail: ark@astro.le.ac.uk}, M.\,E.~Beer,
D.\,J.~Rolfe, K.~Schenker and J.\,M.~Skipp\\ 
Theoretical Astrophysics Group, University of Leicester,
Leicester, LE1~7RH, UK\\
}

\maketitle

\begin{abstract}
We consider the population of black widow pulsars (BWPs). The large
majority of these are members of globular clusters. For minimum
companion masses $<
$\,0.1\,M$_{\odot}$, adiabatic evolution and
consequent mass loss under gravitational radiation appear to provide
a coherent explanation of all observable properties. We suggest that
the group of BWPs with minimum companion masses $\ga
0.1\,\rm{M}_{\odot}$ are systems relaxing to equilibrium after a
relatively recent capture event. We point out that all binary
millisecond pulsars (MSPs) with orbital periods $P \la 10$\,hr are
BWPs (our line of sight allows us to see the eclipses in 10 out of 16
cases). This implies that recycled MSPs emit either in a wide fan beam
or a pencil beam close to the spin plane. Simple evolutionary ideas
favour a fan beam.
\end{abstract}

\begin{keywords}
binaries: close
\end{keywords}

\section{Introduction}

It is widely believed that most millisecond pulsars (MSPs) are spun up
by accretion from a close binary companion. This recycling
(Radhakrishnan \& Srinivasan, 1981) can only take place when the
neutron star magnetic field has decayed to a value $\sim 10^8$~G. If
accretion subsequently stops for some reason, the neutron star appears
as a millisecond pulsar with a very low spindown rate, as its low
magnetic field makes dipole radiation very weak.

In line with these ideas a large proportion of millisecond pulsars are
members of binary systems. Often these pulsars undergo very wide
eclipses, with an obscuring object much larger than the companion
star's Roche lobe. This must be an intense wind from the companion
star, driven in some way by the pulsar emission (Fruchter et al.,
1988). Whenever such eclipses are seen the binary eccentricity and the
pulsar mass function are extremely small, implying companion masses
$M_2 \la \msun$. There is another group of binary millisecond pulsars
with systematically lower mass functions, and it is natural to assume
that these are also evaporating systems with orbital inclinations
which prevent us seeing the eclipses (Freire et al., 2001).

In a recent paper (King et al. 2003; hereafter KDB) we pointed out
that the incidence of black widow pulsars (BWPs) is far higher in
globular clusters than in the field. We identified a favoured
formation mechanism for BWPs in globulars in which turnoff--mass stars
exchange into wide binaries containing recycled millisecond pulsars
(MSPs) and eject their helium white dwarf companions. Once angular
momentum loss or nuclear expansion bring the companion into contact
with its Roche lobe the pulsar is able to expel the matter issuing
through the inner Lagrange point $L_1$. Thus mass is lost on the
binary evolution timescale. BWPs are observable only when this
evolution and thus the mass loss are slow.

Here we consider the consequences of this picture for the subsequent
evolution of BWPs. Our main aim is to understand the distribution of
BWPs in the plane of minimum companion mass $M_{\rm min}$ and
orbital period $P$ (Fig.~\ref{figmsp}). 

\section{Population of Low--Mass BWPs}

\begin{figure}
\includegraphics[width=8.4cm]{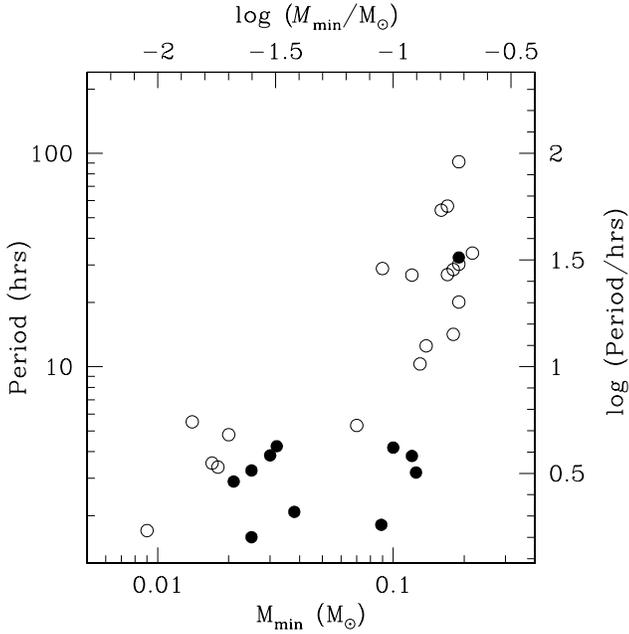}
\caption{A plot of log period in hours versus log minimum companion
  mass for all the binary MSPs in globulars. Filled circles denote 
  eclipsing systems and open circles non-eclipsing systems.}
\label{figmsp}
\end{figure}

BWPs appear to fall into two distinct groups which we shall call
high--mass and low--mass, depending on whether $M_{\rm min} \ga $ or
$\ga 0.1\msun$. The high--mass group have noticeably more
irregular eclipses than the low--mass group (e.g. Scott Ransom, talk
at the Aspen Meeting on Binary Millisecond Pulsars in January
2004). In the standard way we may associate with these two groups
other binary MSPs which do not eclipse but whose values of $M_{\rm
min}$ are even smaller than those of the eclipsing systems. The
interpretation is that these are BWPs seen at low orbital
inclination. With this association the distribution of
Fig.~\ref{figmsp} puts the dividing line between high-- and low--mass
BWPs at $M_{\rm min} \simeq 0.05\,\rm{M}_{\odot}$.

\begin{figure}
\includegraphics[width=8.4cm]{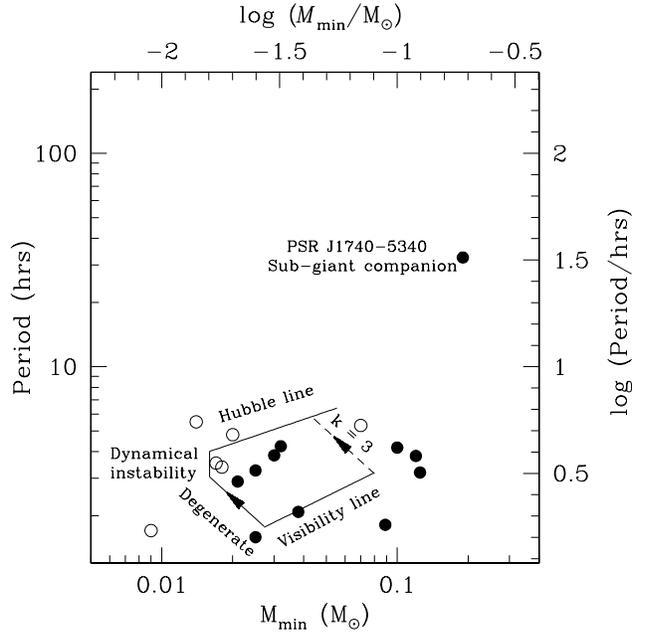}
\caption{A plot of log period in hours versus log minimum companion
  mass for all the BWPs in globulars along with the evolutionary
  constraints as described in the text for the low-mass systems. Also
  labelled is the longest period BWP PSR J1740--5340 which has a
  sub-giant companion.}
\label{figbwp}
\end{figure}

We consider the low--mass group first. The small
secondary mass (close to $M_{\rm min}$ for eclipsing systems) implies
a thermal time much longer than the binary evolution timescale for mass
loss causing BWP behaviour. The star therefore reacts
adiabatically. As it is either fully convective or degenerate we can
model approximately it as an $n=3/2$ polytrope, with radius
\begin{equation}
R_2 \simeq 10^9(1+X)^{5/3}m_2^{-1/3}k ~{\rm cm}
\label{r2}
\end{equation}
(cf e.g. King 1988; for more accurate representations see Nelson \&
Rappaport, 2003). Here $m_2 = M_2/\rm{M}_{\odot}, X$ is the
fractional hydrogen content, and $k \geq 1$ measures the deviation
from the fully--degenerate radius given by $k = 1$. We assume that
low--mass BWPs have a range $1 < k \la$ few for reasons we will
discuss in the next section.  
Assuming a Roche-lobe filling donor gives the mass-period relation
\begin{equation}
m_2 = 1.5 \times 10^{-2} (1+X)^{5/2} k^{3/2} P_{\rm h}^{-1}
\label{mp}
\end{equation}
where $P_{\rm h}$ is the orbital period measured in hours. We assume
further that loss of orbital angular momentum via gravitational
radiation (GR) drives orbital evolution, so that
\begin{equation}
{\dot M_2\over M_2} = {3(1+q)\over 2 - 3\beta + q}{\dot J_{\rm
    GR}\over J}
\label{dotm}
\end{equation}
where $q=M_2/M_1$, $M_1 \simeq
1.4\,\rm{M}_{\odot}$ is the pulsar mass, and 
$\beta$ is the specific angular momentum of the lost mass
relative to that of the secondary (van Teeseling
\& King 1998). Thus

\begin{equation}
\beta = \left ( \frac{b}{a} \right )^2 (1+q)^2
\end{equation} 
where $a$ is the binary separation and $b$ is the distance from the
centre of mass at which the matter is ejected from the binary. This
is between the circularization radius of the infalling matter and
the $L_1$ point.  As all BWPs have $M_2 \ll M_1$, $\beta \ll 2/3$
except for very small $q$ (see below) the first term on the right hand
side of (\ref{dotm}) is $3/2$ and we can rewrite (\ref{dotm}) as
\begin{equation}
\dot M_2 = -1.9 \times 10^{-10}\,P_{\rm
  h}^{-8/3}\,m^{2/3}\,m_{0.1}^2~\rm{M}_{\odot}{\rm yr}^{-1}
\label{dotm2}
\end{equation}
where $m = (M_1 + M_2)/\rm{M}_{\odot}$ and $m_{0.1} = m_2/0.1$. According to
KDB the system is only visible as a BWP if this rate is less than the
critical value 
\begin{equation}
\dot M_{\rm crit} = 1.5\times 10^{-12}\,P_{\rm h}\,
T_6^{3/4}~\rm{M}_{\odot}{\rm yr}^{-1}
\label{crit}
\end{equation}
for free--free absorption at typical (400 -- 1700 MHz) observing
frequencies, where $T_6 \sim 1$ is the temperature of the lost mass
near $L_1$ in units of $10^6$~K. The resulting constraint
\begin{equation}
P_{\rm h} \ga 3.7 \,T_6^{-9/44}\, m^{2/11}\, m_{0.1}^{6/11}
\label{vis}
\end{equation}
is plotted on Fig.~\ref{figbwp} as `Visibility line'.

In a similar way we can plot various other constraints on this
figure. For very small mass ratios $q = M_2/M_1$ the $\beta$ term in the
denominator of (3)  becomes significant. This signals dynamical
instability, as the Roche lobe moves inwards wrt the stellar surface
(cf Stevens, Rees \& Podsiadlowski, 1992). At such masses the
companion must be broken up and sheared into a large disc surrounding
the pulsar. This is presumably the origin of the planets observed
around PSR~B1257+12 (Wolszczan \& Frail 1992; Konacki \&
Wolszczan 2003). Direct numerical calculation of $b$ from
Roche geometry shows that dynamical instability occurs when $M_2/M_1 <
0.02$ if matter is ejected from the $L1$ point. However, matter can be
ejected from anywhere between the $L1$ point and the circularization
radius (which even for small mass ratios is not comparable with
the distance to the $L1$ point). If matter
reaches in 5 per cent of the distance from the centre of mass to the
$L1$ point this gives $M_2/M_1=0.011$ i.e. $m_2 = 0.016$. This line is
labelled `Dynamical Instability' on Fig.~\ref{figbwp}. We also plot 
the line (`Hubble line') on which the binary evolution has a
characteristic timescale longer than a Hubble time. Finally we plot
the binary evolution of a system whose secondary is fully degenerate
(`Degenerate') and one whose radius is 3 times as large for the same
mass ($k = 3$) as given by equation (\ref{mp}). Arrows on the degenerate
sequences indicate the direction of evolution.

If this is a viable picture of the evolution of low--mass BWPs, we
expect them to have combinations of $P, M_2$ lying inside
the various constraints shown in Fig.~\ref{figbwp}. Given that
eclipsing systems (solid circles) have $M_2 \simeq M_{\rm min}$, while
non--eclipsing systems have $M_2$\,$>$\,$M_{\rm min}$, we see that the
data plotted on Fig.~\ref{figbwp} are indeed reasonably consistent
with adiabatic evolution under mass loss driven by gravitational
radiation.

\section{High--Mass BWPs}

We now consider the remaining systems on Fig.~\ref{figbwp}, i.e. those with
$M_{\rm min} \ga 0.05\,\rm{M}_{\odot}$. Nelson (talk at the
Aspen Meeting on Binary Millisecond Pulsars in January 2004) has
studied the evolution of the long--period system J1740--5340, which
has a subgiant companion (D'Amico et al. 2001; Ferraro et al. 2001)
and shown that it is consistent with a location near a bifurcation
point, at which nuclear evolution and angular momentum loss are
comparable. The remaining 4 systems have $P \la 5$\,hr and $M_{\rm
min} \la 0.13\,\rm{M}_{\odot}$. Except for implausibly small orbital
inclinations these companions have main--sequence radii far too
small to fill their Roche lobes. There are various ways around this
difficulty, including competition between nuclear and orbital
evolution, or attributing the eclipses to strong stellar winds of
detached companions. However the most likely explanation appears to us
to follow from the turbulent history these binaries must have had. 

The process of gaining a new companion must involve considerable
disturbance to that star, probably leading to extensive mass
loss. This is clearly indicated for tidal capture (Podsiadlowski
1996), but is probably true in any picture. For an exchange encounter
to give a short period system a large post-exchange eccentricity is
required (see section~\ref{discuss}). When this eccentric system
circularizes the tidal effects on the secondary can in some cases be
similar to tidal capture if the periastron distance is 
comparable to the stellar radii. The globular--cluster 
X--ray binary AC211 (van Zyl et al. 2004; Charles, Clarkson \& van Zyl
2002, and references therein) may be an example 
where the companion star is oversized because of the capture process,
the only difference being that the neutron star in this system
accretes the overflowing matter rather than expelling it, presumably
because no previous partner recycled it. For an exchange encounter the
lowest mass star is ejected and so the captured star, in a high-mass
BWP formed through this route, must have lost a significant portion of
its initial mass. This mass loss, however, occurs during the tidal
circularization (not afterwards) and so the companion could still be
thermally bloated. 

After the tidal disturbance and consequent mass loss, the companion 
attempts to reach its new main--sequence radius on a thermal
timescale. This process competes with orbital shrinkage via
GR and reduces the mass loss (already weak -- cf
equation (\ref{dotm2}) for the relatively long orbital periods of most
of this group). It is therefore plausible that BWPs ultimately emerge
from the tidally induced mass loss with the parameters of the
high--mass BWPs. The thermal timescale of some of these BWPs may be
less than the GR timescale and so eventually these would shrink from
contact with their Roche lobes. However, they would still be observed
as high-mass BWPs for the length of their thermal timescale.

If the companion has a sufficiently strong 
stellar wind, this can produce orbital eclipses through
free--free absorption. This appears to happen in PSR 1718--19 (Wijers
\& Paczy\'nski 1993; Burderi \& King 1994). Here $P \simeq 6$\,hr, and
modelling of the absorption light curve (Burderi \& King 1994) shows
that $M_2 \sim 0.2\,\rm{M}_{\odot}$. We note that PSR 1718--19 is
probably a cluster member (Wijers \& Paczy\'nski 1993) and that the
stellar wind must be the eclipsing agent as the pulsar is not an MSP,
and thus incapable of driving mass loss.

Although the companion must relax towards the main sequence, its 
radius ultimately shrinks only slowly, whereas orbital shrinkage via
GR accelerates. Depending on the initial separation, the binary
reaches contact with the companion somewhat oversized compared with
its thermal--equilibrium radius. This is probably the origin of the
range of radii ($k$--values) inferred for low--mass BWPs above.

Grindlay (talk at Aspen Meeting on Binary Millisecond Pulsars in
January 2004) has found 108 X-ray sources in the globular cluster 47
Tuc. A number of these are claimed to be quiescent low-mass X-ray
binaries (LMXBs) and have measured periods from X-ray dips/eclipses
(eg W37 at 3\,hrs), power law components (possibly from a wind) and
variable absorption. These may be BWPs, rather than quiescent LMXBs,
where the absorption is too large to observe the radio emission and
the stellar wind is responsible for the X-ray emission.

\section{Fan or Pencil Beam?}

Fig.~\ref{figmsp} reveals the striking fact that {\it all binary MSPs with
  orbital periods} $P \la 10$\,{hr} {\it are BWPs: 10 out of 16
  actually eclipse.} There are two obvious possible explanations for
  this:

(a) Fan beam. The pulsar beam of a recycled MSP is so wide that it
  always includes the orbital plane, whatever the relative orientation
  of spin and orbit. For a wide fan beam such systems are detectable
  at any spin inclination. Then all binary MSPs become BWPs as soon as
  their companions fill their Roche lobes.

(b) Pencil beam. The beam of a recycled MSP is narrowly
  confined. Clearly the only plausible geometry for making BWPs has
  the beam axis orthogonal to the spin, with the latter roughly
  aligned with the binary orbit. 

It is quite difficult to break the degeneracy between these two
possibilities. However the pencil beam requires alignment, and thus
that the pulsar has accreted $\ga 0.1\,\rm{M}_{\odot}$ from its {\it
current} companion. This requires it to have been an LMXB and then
somehow broken contact, and seems harder to reconcile with the picture of
high--mass BWP evolution we have sketched above. We thus tentatively
conclude that a fan beam offers a simple explanation for the
universality of eclipsing behaviour in MSPs with short orbital
periods.

\section{Discussion} \label{discuss}

It is generally agreed that dynamical encounters must be invoked to
explain the overproduction of BWPs in globular clusters (KDB; Rasio et
al. 2000). Rappaport, Putney \& Verbunt (1989) considered the problem
of exchanging out the white dwarf companion of the neutron star in a
binary MSP. Their Fig.~1 shows that the timescale (in 47 Tuc) for
ejecting the white dwarf from systems with periods of 10, 100 and 1000
days is $\sim$10, 3, and 1 Gyr, respectively. Thus an exchange
encounter is likely only for orbital periods $\ga 30$ days.
The post--exchange period is generally not substantially shorter, and
most of these systems would not become BWPs. However the
post--exchange eccentricity $e$ is evenly distributed in $e^2$ (Heggie
1975). For large $e^2$ tides circularize the orbit at much tighter
separations than the initial post-exchange value.  The post-exchange
period after tidal circularization can thus be much shorter (hours) in
a few cases, producing a BWP.

The need for the exchange into the binary MSP is set by the assumption
that radio pulsars do not turn on as long as there is a binary
companion in near contact; thus the need for removing it altogether
after the initial recycling. Burgay et al. (2003), however, claim that
the lack of detection of radio pulsations from quiescent LMXBs could
result from pulsar radiation driving a large outflow from the system
which then attenuates the radio emission. A problem with this scenario
is that the pulsar must remove the accretion disc before it can blow
away matter flowing through the inner Lagrangian point. Burderi et
al. (2001) show that this can only occur in some wide systems, and
only then when the magnetic pressure due to the neutron star overcomes
the disc's internal pressure. Consequently it is unlikely that the
companion in a BWP is the one that spun the pulsar up, unless the
latter has somehow contracted well inside its Roche lobe. If this
occurs through evolutionary loss of the envelope one then has the
problem of bringing the system to short periods in a reasonable time.

Converting wide orbits to narrow ones was the main motivation for
Rasio et al. (2000) in considering an alternative scenario. This
invokes binary formation through exchange interactions, producing
neutron star systems with companions which are more massive. This
must occur in a relatively short interval of $1 - 2$~Gyr before all
the potential extended companions with masses larger than that of the
neutron star have finished their evolution. Given such a system, a
common envelope phase results once the companion evolves and overfills
its Roche lobe. The envelope of the red giant is ejected and the core
spirals in, leaving a He--rich white dwarf in a relatively close orbit
with the neutron star. If contact is established, the pulsar could
presumably begin to evaporate the companion. However because the donor
star has no hydrogen, the orbital periods of systems made this way are
much shorter than those of observed BWPs (see equation \ref{r2} and
Fig.~\ref{figbwp}).

The evolutionary outcome for BWPs depends on whether they have evolved
to a period where they have become dynamically unstable. A number of
BWPs exist which have reached or are evolving towards the `Hubble
line' and so are not yet old enough to have been dynamically
disrupted. The systems which have undergone dynamical instability are
observed as isolated MSPs. In the globular 47 Tuc 7 out of 21 of MSPs
are isolated (see
\href{http://www.naic.edu/~pfreire/GCpsr.html}
{http://www.naic.edu/$\sim$pfreire/GCpsr.html}
for an up to date list of pulsars in globular clusters) while 5 out of
21 are low-mass BWPs. Objects may undergo evolution through the
low-mass evolutionary route and disruption of the companion. However,
the number of resulting isolated MSPs should not be significantly
greater than the number of BWPs unless there is a process which
strongly favours short periods (less bloated secondaries). The similar
number of low-mass BWPs and isolated pulsars in 47 Tuc supports this
conclusion.

\section*{Acknowledgements} 

We thank the referee Saul Rappaport for comments which have improved
this paper. We are also grateful to the organisers of the Aspen 
Meeting on Binary Millisecond Pulsars in January 2004 which provided
much stimulus for this work. DR and KS are supported by the Leicester
PPARC rolling grant for theoretical astrophysics, and JS by a PPARC
studentship. ARK gratefully acknowledges a Royal Society Wolfson
Research Merit Award. MEB acknowledges the support of a UKAFF
fellowship.

{\vspace{0.5cm}\small\noindent This paper
has been typeset from a \TeX / \LaTeX\ file prepared by the author.}

\label{lastpage}

\end{document}